# Interchangeability of Combined Piezoelectrooptic Effect in LiTaO$_3$ and LiNbO$_3$ Crystals

Mys O., Vlokh R.

Institute of Physical Optics, 23 Dragomanov St., 79005 Lviv, Ukraine
e-mail: vlokh@ifo.lviv.ua

## Abstract

In the present paper the coefficients of piezooptic (PO), electrooptic (EO) and combined piezoelectrooptic (PEO) effects for LiNbO$_3$ and LiTaO$_3$ crystals are reported. It is shown that the combined PEO effect is interchangeable, i.e. its coefficients determined independently from the EO and PO experiments are the same.

## Introduction

The combined piezoelectrooptic (PEO) effect consists in changes of optical-frequency dielectric impermeability constants $B_{ij}$ (or the refractive indices $n$, where $B_{ij} = \left(\frac{1}{n^2}\right)_{ij}$) of crystals under a joint influence of electric field (the components $E_m$) and mechanical stress (the components $\sigma_{kl}$) [1]. It is described by the relation

$$\Delta B_{ij} = N_{ijklm}\sigma_{kl}E_m + M_{ijklmn}\sigma_{kl}E_m E_n, \quad (1)$$

where $N_{ijklm}$ is a fifth-rank polar tensor with the internal symmetry $[V^2]^2 V$ and $M_{ijklmn}$ polar tensor of a rank six with the symmetry $[V^2]^3$. It follows from the internal symmetry of $N_{ijklm}$ that the PEO effect is linear in the electric field and can exist only in non-centrosymmetric crystals. This effect may be also presented as a change in EO coefficients under the action of mechanical stress:

$$\Delta r_{ijm} = N_{ijmkl}\sigma_{kl}, \quad (2)$$

or, equivalently, as a change in PO coefficients under the action of the electric field:

$$\Delta \pi_{ijkl} = N_{ijklm}E_m. \quad (3)$$

Thus, the components of the tensor $N_{ijklm}$ may be derived while measuring the changes in the EO coefficients under the action of mechanical stress, as well as from the changes in the PO coefficients under the action of electric field:

$$N_{ijklm} = \frac{\Delta \pi_{ijkl}}{E_m} = \frac{\Delta r_{ijm}}{\sigma_{kl}}. \quad (4)$$

From this point of view, the combined PEO effect may be called as interchangeable.

In our previous paper related to studies of interchangeability of the combined PEO effect in LiTaO$_3$ [2], an error has occurred in the processing of results and their presentation. Due to the calculation error, the presented results concerned with the birefringence induced by the electric field and the mechanical stresses have in fact proved to be two times smaller.

In this paper, we present the results of reinvestigation of the combined PEO effect in LiTaO$_3$ crystals. We show experimentally, on the example of LiTaO$_3$ and LiNbO$_3$ crystals, that the effect is indeed interchangeable.

## Experimental

LiNbO$_3$ and LiTaO$_3$ single crystals with plane-parallel faces perpendicular to the principal axes of optical permeability ellipsoid were polished with the diamond powders and paste. He-Ne laser radiation with the wavelength of 632.8 nm propagated along $Y$ direction. The electric field was applied along $Z$ axis, using the copper electrodes, whereas the mechanical stress was applied along $X$ direction. Using the Senarmont technique, we measured the changes of the birefringence due to the electric field, occurring at different magnitudes of the mechanical stress, and vice versa. The optical birefringence increment was calculated with the formula

$$\delta(\Delta n) = \frac{\Delta \varphi \lambda}{\pi d_k}, \qquad (5)$$

where $\Delta \varphi$ is the rotation angle of the polarization plane behind the quarter-wave plate, $\lambda$ the laser radiation wavelength and $d_k$ the thickness of sample along the propagation direction. The total errors of determination of EO and PO coefficients do not exceed 13% and 18%, respectively, while the total error for the PEO coefficients is 26%. These total errors consist of systematic and statistical parts. The systematic one does not affect determination of the relative EO or PO birefringence increments detected at different values of the mechanical stress or electrical field, respectively. However, it has effect on the accuracy of determination of the absolute values of PO, EO and PEO coefficients. The systematic errors appearing during the measurements of EO and PO coefficients are equal respectively to 8% and 12%. On the plots represented below, we indicate only the statistical errors, which are 5% and 6% for the cases of EO and PO measurements, respectively.

**Results and Discussion**

The matrices of the PO and EO tensors for the point symmetry group 3m are presented, e.g., in [3]. We have also derived the matrix of a fifth-rank polar tensor describing the combined PEO effect. It has the following form:

|  | $\sigma_1 E_1$ | $\sigma_2 E_1$ | $\sigma_3 E_1$ | $\sigma_4 E_1$ | $\sigma_5 E_1$ | $\sigma_6 E_1$ | $\sigma_1 E_2$ | $\sigma_2 E_2$ | $\sigma_3 E_2$ | $\sigma_4 E_2$ | $\sigma_5 E_2$ | $\sigma_6 E_2$ | $\sigma_1 E_3$ | $\sigma_2 E_3$ | $\sigma_3 E_3$ | $\sigma_4 E_3$ | $\sigma_5 E_3$ | $\sigma_6 E_3$ |
|---|---|---|---|---|---|---|---|---|---|---|---|---|---|---|---|---|---|---|
| $\Delta B_1$ | 0 | 0 | 0 | 0 | $N_{151}$ | 0 | 0 | 0 | $N_{132}$ | $N_{151}$ | 0 | 0 | $N_{113}$ | $-N_{113}$ | $N_{133}$ | $N_{143}$ | 0 | 0 |
| $\Delta B_2$ | 0 | 0 | 0 | 0 | $-N_{151}$ | 0 | 0 | 0 | $-N_{132}$ | $-N_{151}$ | 0 | 0 | $N_{113}$ | $-N_{113}$ | $N_{133}$ | $-N_{143}$ | 0 | 0 |
| $\Delta B_3$ | 0 | 0 | 0 | 0 | $N_{351}$ | $N_{312}$ | $-N_{312}$ | $N_{312}$ | 0 | $N_{351}$ | 0 | $-N_{311}$ | $N_{313}$ | $N_{313}$ | $N_{333}$ | 0 | 0 | 0 |
| $\Delta B_4$ | 0 | 0 | 0 | 0 | $N_{451}$ | 0 | 0 | 0 | $N_{531}$ | $-N_{451}$ | 0 | 0 | $N_{413}$ | $-N_{413}$ | 0 | $N_{443}$ | 0 | 0 |
| $\Delta B_5$ | 0 | 0 | $N_{531}$ | $N_{451}$ | 0 | 0 | 0 | 0 | 0 | 0 | $-N_{451}$ | 0 | 0 | 0 | 0 | 0 | $N_{443}$ | $N_{413}$ |
| $\Delta B_6$ | 0 | 0 | $N_{132}$ | $N_{151}$ | 0 | 0 | 0 | 0 | 0 | $N_{151}$ | 0 | 0 | 0 | 0 | 0 | 0 | $N_{143}$ | $N_{113}$ |

According to the appearance of the EO, PO and the combined PEO tensors, the equation of optical indicatrix, perturbed by the applied electric field $E_3$ and mechanical stress $\sigma_1$, for the crystals belonging to the symmetry group 3m may be written as

$$(B_1 + r_{13}E_3 + \pi_{11}\sigma_1 + N_{113}\sigma_1 E_3)X^2 + (B_1 + r_{13}E_3 + \pi_{21}\sigma_1 + N_{113}\sigma_1 E_3)Y^2 + \\ (B_3 + r_{33}E_3 + \pi_{31}\sigma_1 + N_{313}\sigma_1 E_3)Z^2 + 2\pi_{41}\sigma_1 ZY + 2N_{413}\sigma_1 E_3 ZY = 1 \qquad (6)$$

When we neglect the optical indicatrix rotation (the estimated indicatrix rotation angle does not exceed 10' for the mechanical stresses applied by us), the total change of the birefringence becomes

$$\delta(\Delta n)_{zx} = \delta(\Delta n)_{zx}^E + \delta(\Delta n)_{zx}^\sigma + \delta(\Delta n)_{zx}^{E\sigma} = \\ \frac{1}{2}\Big[(n_3^3 r_{33} - n_1^3 r_{13})E_3 + (n_3^3 \pi_{31} - n_1^3 \pi_{11})\sigma_1 + (n_3^3 N_{313} - n_1^3 N_{113})\sigma_1 E_3\Big], \qquad (7)$$

where $\delta(\Delta n)_{zx}^E$, $\delta(\Delta n)_{zx}^\sigma$ and $\delta(\Delta n)_{zx}^{E\sigma}$ are respectively the birefringences induced by the electric field and the mechanical stress alone and the two external factors acting simultaneously.

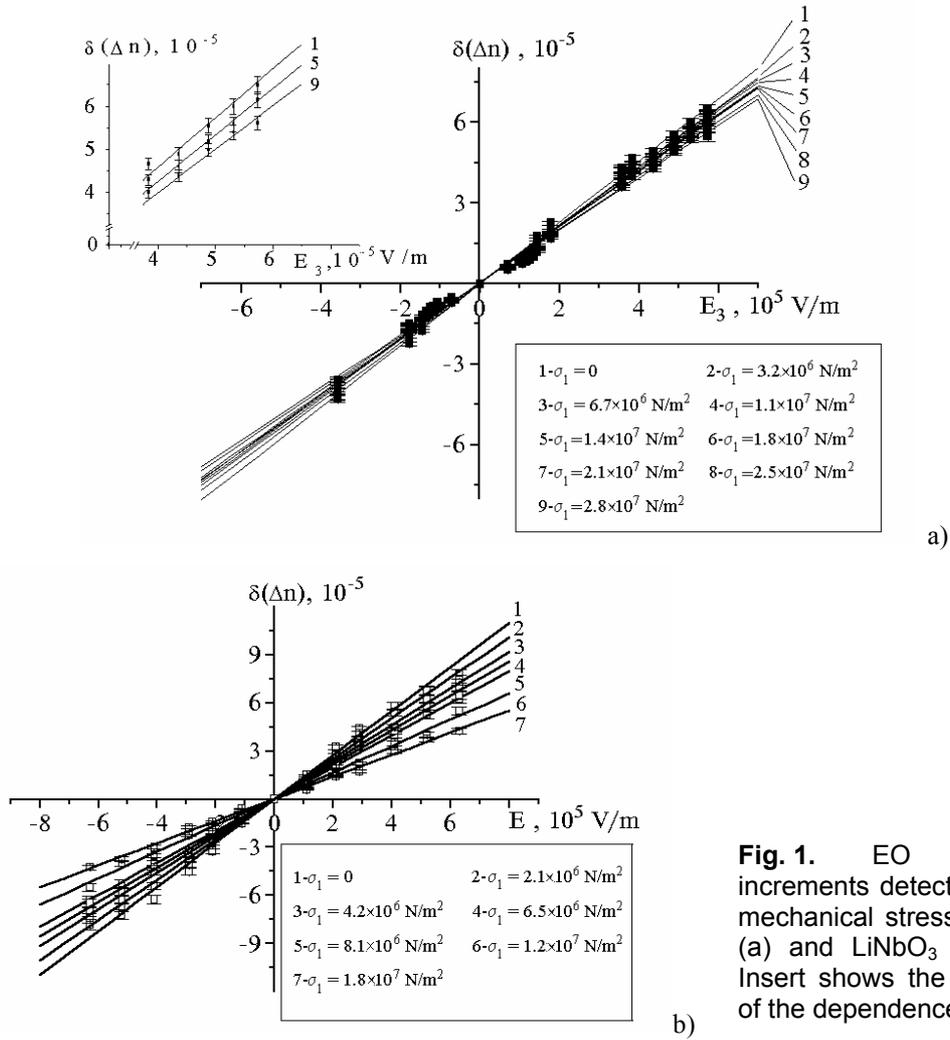

**Fig. 1.** EO birefringence increments detected at different mechanical stresses for LiTaO$_3$ (a) and LiNbO$_3$ (b) crystals. Insert shows the enlarged part of the dependence.

As a consequence, the coefficients of the EO, PO and the combined PEO effects can be calculated according to the formulae

$$n_3^3 r_{33} - n_1^3 r_{13} = \frac{2\delta(\Delta n)_{zx}^E}{E_3}, \quad (8)$$

$$n_3^3 \pi_{31} - n_1^3 \pi_{11} = \frac{2\delta(\Delta n)_{zx}^\sigma}{\sigma_1}, \quad (9)$$

$$n_3^3 N_{313} - n_1^3 N_{113} = \frac{2\delta(\Delta n)_{zx}^{\sigma E}}{\sigma_1 E_3}. \quad (10)$$

The changes in the EO birefringence increment for LiTaO$_3$ and LiNbO$_3$ crystals that take place at different mechanical stresses are presented in Fig. 1. The similar changes in the PO birefringence increment at different electric fields are presented in Fig. 2. Using Eqs. (8) and (9), one can calculate the dependence of EO coefficient on the mechanical stress and that of PO coefficient on the electric field (see Fig. 3).

On the basis of Eqs. (2) and (3) and the above dependences, we are able to calculate the coefficient of the combined PEO effect. It has been found that the relevant coefficient determined from the dependence of EO coefficients on the mechanical stress is equal to $n_3^3 N_{313} - n_1^3 N_{113} = (6.2 \pm 1.6) \times 10^{-18} m^3/NV$ for LiNbO$_3$ crystals and $n_3^3 N_{313} - n_1^3 N_{113} = (10.7 \pm 2.8) \times 10^{-19} m^3/NV$ for LiTaO$_3$ crystals. The same PEO coefficient derived from the dependence of PO coefficients on the electric field is $n_3^3 N_{313} - n_1^3 N_{113} = (4.6 \pm 1.2) \times 10^{-18} m^3/NV$ for LiNbO$_3$ and $n_3^3 N_{313} - n_1^3 N_{113} = (8.6 \pm 2.2) \times 10^{-19} m^3/NV$ for LiTaO$_3$. One can see that the two PEO coeffi-

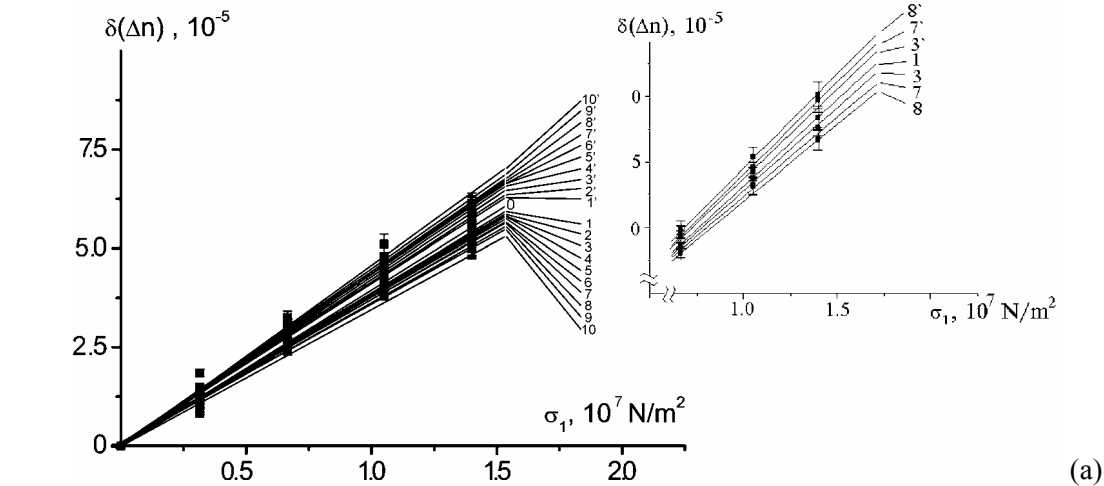

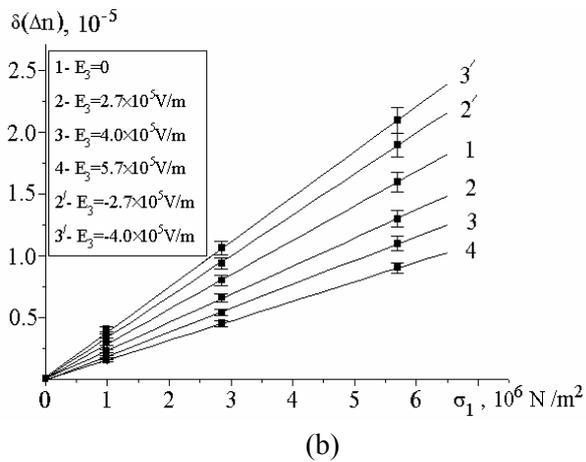

**Fig. 2.** PO birefringence increments detected at different electric fields for LiTaO$_3$ (a) (curve 0 – $E_3$=0; 1– 3.8x10$^5$; 2– 4.7x10$^5$; 3– 5.1x10$^5$; 4– 5.3x10$^5$; 5– 5.7x10$^5$; 6– 6.5x10$^5$; 7– 7.4x10$^5$; 8– 8.2x10$^5$; 9– 8.9x10$^5$; 10– 9.6x10$^5$ and 1'– -3.8x10$^5$; 2'– -4.7x10$^5$; 3'– -5.1x10$^5$; 4'– -5.3x10$^5$; 5'– -5.7x10$^5$; 6'– -6.5x10$^5$; 7'– -7.4x10$^5$; 8'– -8.2x10$^5$; 9'– -8.9x10$^5$; 10'– -9.6x10$^5$ V/m, insert shows the enlarged part of the dependence) and LiNbO$_3$ (b) crystals.

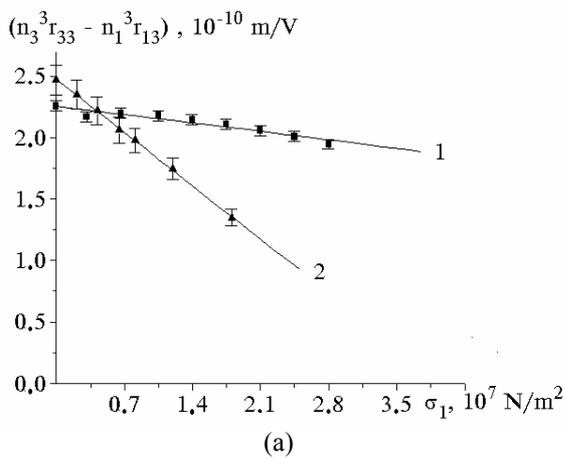
(a)

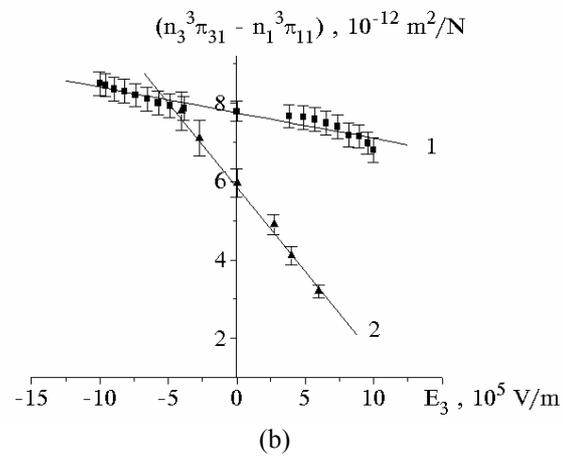
(b)

**Fig. 3.** Dependences of EO coefficient on the mechanical stress (a) and PO coefficient on the electric field (b) for LiTaO$_3$ (1) and LiNbO$_3$ (2) crystals.

cients obtained from both the EO and PO experiments are the same, if one accounts for the experimental errors.

## Conclusions

1. The coefficients of the PO, EO and the combined PEO effects in LiNbO$_3$ and LiTaO$_3$ crystals have been determined for $\lambda$=632.8 nm and $T$ = 291 K. They are as follows:

$$n_3^3 r_{33} - n_1^3 r_{13} = (2.6 \pm 0.3) \times 10^{-10} m/V$$ at $\sigma_1 = 0\ N/m^2$ for LiNbO$_3$ crystals (for a comparison, $n_3^3 r_{33} - n_1^3 r_{13} = 2.24 \times 10^{-10} m/V$, according to [4]);

$n_3^3 r_{33} - n_1^3 r_{13} = (2.3 \pm 0.3) \times 10^{-10} \, m/V$ at $\sigma_1 = 0 \, N/m^2$ for LiTaO$_3$ crystals (the same value is presented in [5]);

$n_3^3 \pi_{31} - n_1^3 \pi_{11} = (5.8 \pm 1.0) \times 10^{-12} \, m^2/N$ at $E_3 = 0 \, V/m$ for LiNbO$_3$ crystals (for a comparison, $n_3^3 \pi_{31} - n_1^3 \pi_{11} = 7.2 \times 10^{-12} \, m^2/N$, according to [5]);

$n_3^3 \pi_{31} - n_1^3 \pi_{11} = (7.8 \pm 1.4) \times 10^{-12} \, m^2/N$ at $E_3 = 0 \, V/m$ for LiTaO$_3$ crystals (for a comparison, $n_3^3 \pi_{31} - n_1^3 \pi_{11} = 10.3 \times 10^{-12} \, m^2/N$, according to [6]);

$n_3^3 N_{313} - n_1^3 N_{113} = (5.4 \pm 1.4) \times 10^{-18} \, m^3/NV$ for LiNbO$_3$ crystals (the mean value);

$n_3^3 N_{313} - n_1^3 N_{113} = (9.6 \pm 2.5) \times 10^{-19} \, m^3/NV$ for LiTaO$_3$ crystals (the mean value).

2. It has been shown that the combined PEO effect is interchangeable, i.e. the coefficients of this effect determined from the EO and PO experiments are the same.

**Acknowledgement**

The authors acknowledge financial support from the Ministry of Education and Science of Ukraine (Project No 0103U000701).